# New strategy to promote conversion efficiency using high-index nanostructures in thin-film solar cells

DongLin Wang[1] & Gang Su[1,*]

[1]Theoretical Condensed Matter Physics and Computational Materials Physics Laboratory, School of Physics, University of Chinese Academy of Sciences, P. O. Box 4588, Beijing 100049, China.
*Correspondence and requests for materials should be addressed to G.S. (email: gsu@ucas.ac.cn).

**Abstract:** Nano-scaled metallic or dielectric structures may provide various ways to trap light into thin-film solar cells for improving the conversion efficiency. In most schemes, the textured active layers are involved into light trapping structures that can provide perfect optical benefits but also bring undesirable degradation of electrical performance. Here we propose a novel approach to design high-performance thin-film solar cells. In our strategy, a flat active layer is adopted for avoiding electrical degradation, and an optimization algorithm is applied to seek for an optimized light trapping structure for the best optical benefit. As an example, we show that the efficiency of a flat a-Si:H thin-film solar cell can be promoted close to the certified highest value. It is also pointed out that, by choosing appropriate dielectric materials with high refractive index (>3) and high transmissivity in wavelength region of 350nm-800nm, the conversion efficiency of solar cells can be further enhanced.

Solar radiation is a clean, stable and abundant energy source. The photovoltaic (PV) device with high efficiency and low cost is essential for extensive utilization of solar energy. Today, the PV market is dominated by wafer based crystalline Si (c-Si) solar cell, which has the efficiency as high as 22% in commercial modules[1,2]. However, the high cost of existing PV modules is a key stumbling block for widespread use of PV technology. Alternative thin film solar cell may provide some viable ways to realize win-win situation between cost and efficiency[3,4]. A thinner active layer has a number of advantages, such as decreasing the material usage for a lower cost, reducing the carrier collection length and bulk recombination for a higher efficiency[5-7]. However, a key premise behind above benefits is to maintain a high light absorption in the thin active region[8]. Traditional method for enhancing the light absorption in thick solar cells is to employ a macroscopic texture on the front surface, which redirects the light propagation to increase the light path in the active layer[9]. Nonetheless, this macroscopic structure is not suitable for thin film solar cells where the thickness may be only few hundred nanometers. Exploration on an effective light trapping structure appropriate for a thin absorber is therefore urgent and crucial in developing the thin film solar cells with high efficiency and low cost.

Thin film hydrogenated amorphous silicon (a-Si:H) solar cells are potential candidates for the global terawatt range deployment, as it is based on nontoxic, abundant and stable materials[10,11]. Random pyramidal texture has demonstrated a remarkable light trapping capability to achieve the certified highest conversion efficiency (10.1%) in a single junction a-Si:H solar cell[2,12]. Recently, plasmonic based nanostructures that can couple the light into the thin a-Si:H layer[13-20] have received significant attention. In particular, nanodome or nanocone plasmonic back reflectors employed a-Si:H thin film solar cells have achieved the highest short circuit current (17.5mA/cm$^2$)[21], fill factor (~69%)[22] and conversion efficiency (~10%)[23,24] for periodic light trapping structures. Mie resonator in dielectric nanostructures[25] and whispering gallery resonator in spherical nanoshells[26] have also revealed an effective light trapping in a-Si:H thin film solar cells. However, these schemes involve the structured active layers that may bring localized defects in materials. Those defects may increase the carrier recombination and degrade the electrical properties of the solar cells. As a result, the performance of solar cell is a tradeoff between the optical benefits and the electrical degradations[23,24,27,28]. It is better that the ultimate light trapping structures only lead to the enhancement of optical absorption but avoid a direct involvement of the active layers.

Recently, some innovative light trapping structures that separate from the active layer have been proposed, including plasmonic based metallic nanostructures[29-31], patterned transparent conductive oxides[32] and dielectric nanosphere arrays[33,34]. In these schemes, the flat active layers were employed with the best quality of materials. However, the maximal enhancement of the efficiency from light trapping is only 11% in a thin a-Si:H solar cell[34]. The optical performance of those light trapping structures still needs to be optimized and improved. In addition, the full field optical and electrical computer simulations can also provide useful information to design the high performance light trapping structures in thin film solar cells[18,19,28,34-36], which, however, rely on one's physical intuition to predefine the geometry of the light trapping structure, and are hard to handle the geometrical variation in seeking for the optimized results[37]. To remedy this, in this work, an effective geometrical optimization algorithm is combined into a full field solar cell simulation process to search for the optimized light trapping structures. To show the power of this method, as an example, we adopt a flat a-Si:H active layer to avoid local defects so that the best electrical performance of the solar cell can be achieved. High refractive indexed scatterers are also considered as light trapping structures that are periodically arranged on the front surface of the flat solar cell and only provide optical benefits. The method proposed in this paper has great potential to better improve the performance of thin film solar cells.

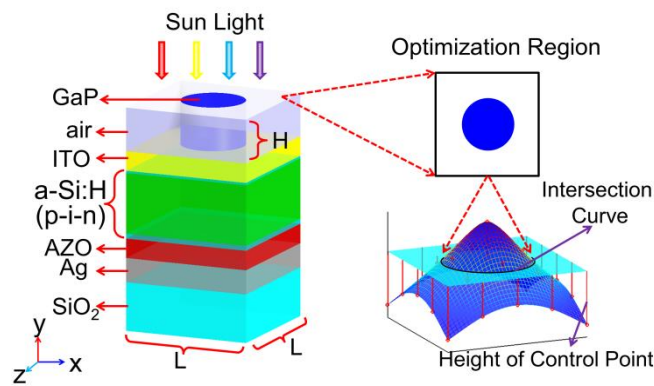

**Figure 1 | Model of an a-Si:H thin film solar cell and schematic depiction of geometry projection method (GPM) used for optimizing the cross section of the dielectric scatterer.** By employing GPM, the cross section shape of the scatterer can be controlled by adjusting the height of the control points. The optimization of the shape can be converted into searching for proper heights of the control points.

## Results

**Tunable light trapping structures based on high-indexed dielectric.** For a-Si:H thin-film solar cells, the ultrathin active layers (<300nm) are necessary for improving the collection efficiency of carriers and reducing the light-induced degradation[38-40]. In a flat solar cell, such a thin absorber can only support several discrete electromagnetic modes. Under a poor light absorption, the effective light trapping can be achieved by coupling the incident light into those guided modes. Usually, there are three ways to implement it, including plasmonic resonator, dielectric resonator and grating diffraction, but which way is the best for light trapping in ultrathin solar cells is still not be established[1]. Recently, high refractive indexed dielectric nanostructures have received much attention, because it can excite optical resonances for trapping the light like metallic nanostructures but has a parasitic absorption lower than metal[25,41,42]. The resonant modes are also sensitive to the shape and the size of the dielectric. It is possible that a proper design can promote those modes to provide the best light trapping in thin film solar cells.

Here we take an example to show the implementation of such a design. As shown in Fig. 1, the periodically arranged GaP scatterers (with high index>3) are placed on the top surface of a flat a-Si:H thin-film solar cell for trapping the light into an active layer. The geometry projection method (GPM)[43,44] is employed to optimize the shape of the dielectric scatterers in seeking for the optimal light trapping structures. This method is more superior to the traditional topology optimization, because it can generate the geometry with smooth edge which is convenient for practical fabrication. The details about the calculation can be found in Method and Supplementary Note 1.

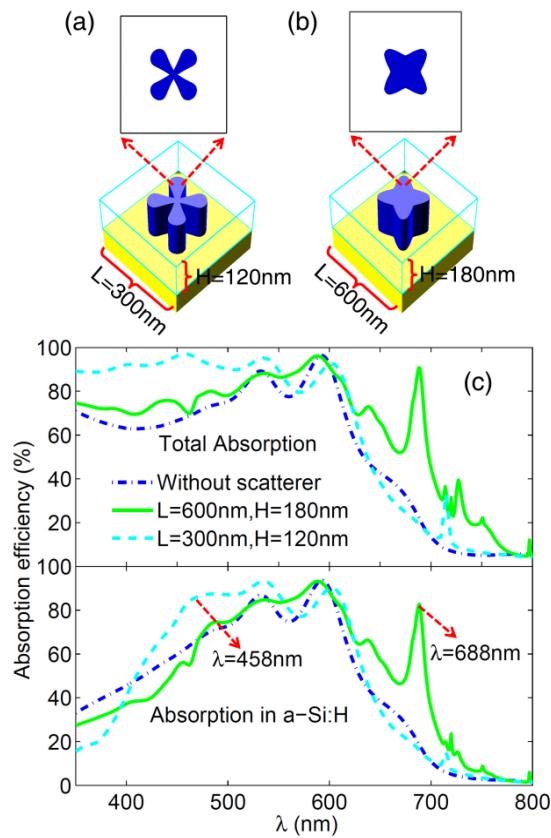

**Figure 2 | Structures of the designed dielectric scatterers and the corresponding light absorption efficiency as function of wavelength in a-Si:H thin-film solar cells.** The optimal dielectric structures for light trapping in the short wavelength region (**a**) and in the long wavelength region (**b**). (**c**) Calculated light absorption efficiency in total solar cell and a-Si:H layer for above two dielectric scatterers. Light absorption efficiency in the flat solar cell without scatterers is shown as a reference.

For the thin film a-Si:H solar cell, most of light can be absorbed in a single light path when the light is with a short wavelength (λ<500nm). In this case, the main loss of light is caused by reflection at the front surface of the solar cell. Here the shape of the cross section of the dielectric scatterer is optimized to reduce the reflection in the short wavelength region (350nm-500nm). As shown in Figure 2a and 2c, the quatrefoil patterned scatterer with height *H*=120nm and patch *L*=300nm demonstrates a perfect antireflection in the short wavelength region. Especially, the total reflection at wavelength λ=458nm can be smaller than 3%. An obvious enhancement of light absorption in the active layer was obtained from λ=400nm to λ=550nm as the result of a remarkable antireflection. However, when λ<400nm the optical loss appears in the active layer due to a large parasitic absorption in GaP scatterer.

In the long wavelength region (λ>600nm), the light absorption is weak in the active layer. The unabsorbed light will be reflected back by the bottom Ag layer, which leads to an insignificant enhancement of the light absorption by the antireflection. In this situation, it is an effective way to couple the incident light into the guided modes in the thin flat active layer to improve the light absorption. Through our optimization calculation, the dielectric scatterer with a different geometrical structure and a perfect light trapping capacity in the long wavelength region (600nm-800nm) was obtained. In Figure 2b and 2c, the four-pointed star shaped scatterer with height $H$=180nm and patch $L$=600nm can improve excellently the light absorption from λ=600nm to λ=800nm. In particular, the absorption in the active layer at wavelength λ=688nm can exceed 80%. In the short wavelength region, however, the parasitic absorption in the scatterer reduces the light absorption in the active layer.

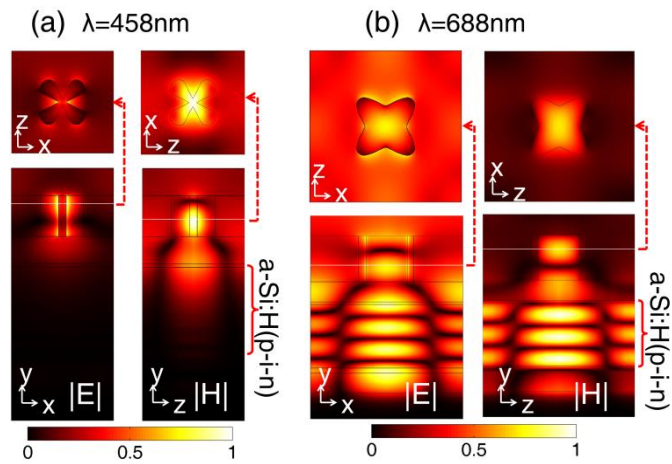

**Figure 3 | Normalized electric and magnetic field intensity profiles inside a-Si:H thin-film solar cells.** The intensity profiles are calculated at the incident light wavelength 458nm for the solar cell with a short wavelength scatterer (**a**) and the wavelength 688nm for the solar cell with a long wavelength scatterer (**b**). The top profiles are perpendicular to the y direction and at the maximal intensity of fields in scatterers. The bottom profiles are selected at the middle of the unit cell, where the intensity of electric fields is shown in the yx plane and the intensity of magnetic fields is in the yz plane.

Figure 3 shows the profiles of the normalized intensity of light field inside the solar cells with the above-designed two scatterers at the wavelength 458nm and 688nm where the best performance appears. At a short wavelength λ=458nm, a proper shaped and sized dielectric excites the optical resonance that can be coupled into the substrate layers to increase the forward scattering and reduce the back reflection (Fig. 3a). This phenomenon is similar to the antireflection coating based on the excitation of a Mie resonance[25]. At a long wavelength λ=688nm, the excited optical resonance can couple the incident light into the waveguide modes, which can increase greatly the light path in the active layer and implement the excellent light trapping (Fig. 3b).

**Flat and stepped scatterers for the optimal light trapping**. To implement a light trapping structure in the whole wavelength region (350nm-800nm), it is necessary to combine the above two kinds of dielectric scatterers in one unit cell. What we expect is that one scatterer can reduce the reflection in the short wavelength region, and another can couple the light into the guided modes in the long wavelength region. To achieve this goal, we design a flat scatterer with an eclectic height $H$=160nm and patch $L$=600nm (Fig. 4a). As is expected, this flat scatterer exhibits the combined capability for improving the total absorption of the solar cell in the whole wavelength region (Fig. 4b). Unfortunately, it is observed that the increase of the size of the scatterer brings more parasitic absorption that reduces the light absorption in the active layer when λ<450nm. More detail about the parasitic absorption in the solar cell can be found in Supplementary Note 2.

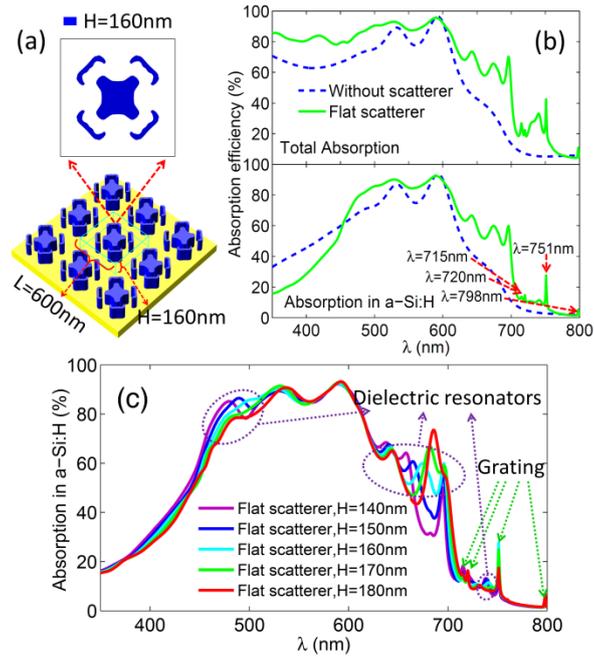

**Figure 4 | Structure of the designed flat dielectric scatterer and the corresponding light absorption as function of wavelength in the a-Si:H thin-film solar cell.** (**a**) The flat dielectric scatterer with height *H*=160nm and patch *L*=600nm has the combined capacity to trap the light in both short and long wavelength regions. (**b**) The calculated light absorption in total solar cell as well as in the a-Si:H layer for the designed flat dielectric scatterer. Four absorption peaks generated by grating appear at wavelength 715nm, 720nm, 751nm and 798nm. Light absorption in the flat solar cell without scatterers is taken as a reference. (**c**) Light absorptions in the a-Si:H layer with the flat scatterer for various heights.

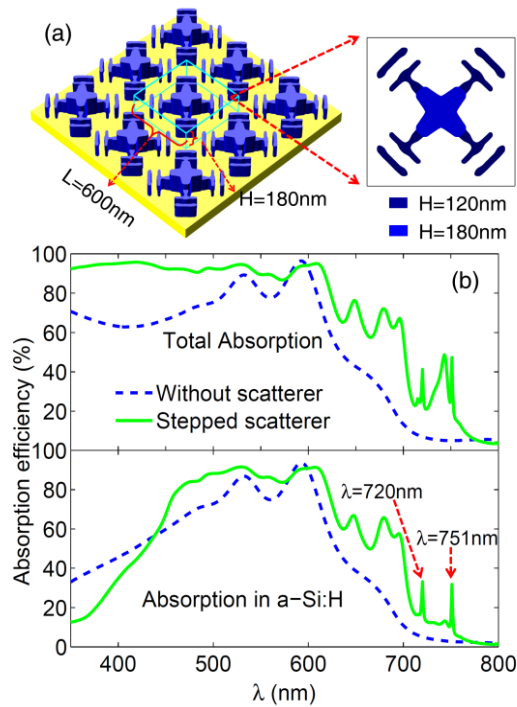

**Figure 5 | Structure of the designed stepped dielectric scatterer and the corresponding light absorption as function of wavelength in the a-Si:H thin-film solar cell.** (**a**) The stepped dielectric scatterer is combined by a lower scatterer with height *H*=120nm and a higher scatterer with height *H*=180nm. (**b**) The calculated light absorption in the total solar cell as well as in the a-Si:H layer for the designed stepped dielectric scatterer. Two absorption peaks generated by grating appear at wavelength 720nm and 751nm. The corresponding light absorption in the flat solar cell without scatterers is taken as a reference.

In this flat scatter, the height (*H*) of the dielectric is an important factor to control the light absorption in the active layer. As shown in Fig. 4c, the light absorption peaks generated by dielectric resonators will offset when the height is changed. The lower scatterer is favorable to reduce the reflection, while the higher scatterer is beneficial for coupling the light into the guided modes. Moreover, there are four light absorption peaks that are insensitive to the height of the scatterer, which are caused by the periodical grating that can also couple the light into the guided modes to enhance the light absorption.

In fact, the eclectic height of the flat scatterer somewhat refrains the dielectric resonator from exhibiting the optimal performance of light trapping. To remedy this shortcoming, we propose another strategy that employs the stepped scatterer which contains both a lower (*H*=120nm) and a higher (*H*=180nm) dielectric structures (Fig. 5 a). It turns out that the perfect antireflection is caused by the optical resonance in the lower dielectric, giving rise to the total absorption in the solar cell exceeding 90% in a broad wavelength region (350nm-550nm) (Fig.5 b). Meanwhile, the optical resonance in the higher dielectric is beneficial for effectively coupling the light into the guided modes. Therefore, this stepped scatterer has more advantages to trap the light into the active layer than the flat one.

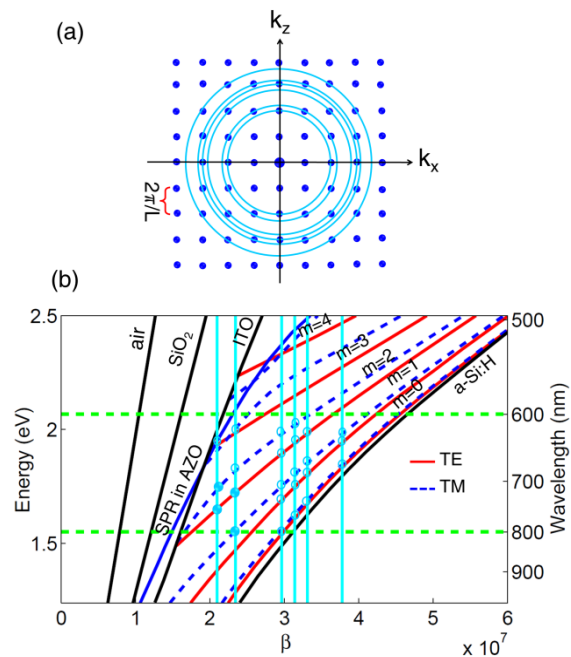

**Figure 6 | Coupling channels in k space and band structures of the guided modes as function of the parallel propagation constant ( $\beta$ ) in the a-Si:H thin-film solar cell.** (**a**) Grating provides several discrete channels (blue dots) that can couple the light into the guided modes in the active layer. The channels intersected by circles represent the eligible match conditions that are available for trapping the light with the wavelength from 600nm to 800nm. (**b**) The calculated band structures of the guided modes for the transverse electric (TE) and the transverse magnetic (TM) polarization in the a-Si:H solar cell. Black and blue solid lines are light cone for air, SiO$_2$, ITO, a-Si:H and the band for the surface plasma resonance (SPR) excited by Ag layer, respectively. The vertical lines correspond to the k values of the eligible coupling channels in (**a**).

**Grating diffraction for light trapping**. The periodically arranged scatterers also have grating effect that can redirect the propagation direction of light through the optical interference and diffraction. In this case, a remarkable enhancement of the light absorption can be achieved when the parallel propagation constant (*β*) of the redirected light and the guided modes can match each other. Figure 6 presents the band structures of waveguide modes as a function of parallel propagation constant for the flat a-Si:H layer. The guided modes with increasing order (m=0, 1, 2, 3, 4) extend from the light cone of a-Si:H to the light cone of indium doped tin oxide (ITO). The bottom Ag layer can excite the surface plasma resonance (SPR) in aluminum doped zinc oxide (AZO) layer, which only has a slight influence on higher order bands of the a-Si:H layer due to the separation by AZO layer. The scatterer in this work can be considered as a two-dimensional square lattice grating with period $L=600$nm. Figure 6a represents several matched channels (blue dots) in $k$ space provided by the grating[10]. The matching condition for the propagation constant can be described by

$$\beta = k_{//} = \sqrt{k_x^2 + k_z^2} \tag{1}$$

For a thin film a-Si:H solar cell, it is more significant to couple the light into the guided modes in the long wavelength region (600nm-800nm). Under the normal incidence, in this region there are six matching conditions indicated by vertical lines in Fig. 6b. (Under the oblique incidence, the match channels of the grating will drift if the incident angle changes). The intersections of the matched lines and the bands of modes suggest that the couplings between the incident light and the guided modes are tenable (Fig. 6). Light couplings caused by grating at the hollow intersections are weaker than those caused by dielectric resonators, so the enhancements of light absorption from grating are merged into those from dielectric resonators. The solid intersections represent the strong grating coupling that leads to a remarkable enhancement for the light absorption in the active layer. Four distinguishable absorption peaks appear in the absorption spectra of the solar cell with the flat scatterer (Fig. 4), while there are two peaks in the absorption spectra of the solar cell with the stepped scatterer (Fig. 5). By contrast, the linewidth of the enhanced absorption spectra from grating is much narrower than that from the dielectric resonator, because the guided modes in the thin active layer and the match channels provided by grating are both discrete.

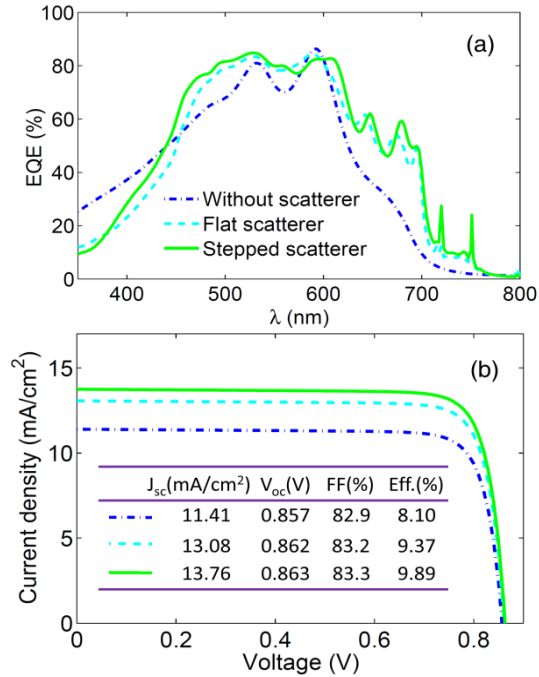

**Figure 7 | External quantum efficiency (EQE) and current-voltage curve for a-Si:H thin-film solar cells with and without scatterers.** (**a**) EQEs for solar cells with and without sactterers. (**b**) current-voltage curve and main performance parameter for solar cells with and without scatterers. The short circuit current density ($J_{sc}$), the open circuit voltage ($V_{oc}$), the fill factor (FF) and the efficiency for solar cells are embedded in (**b**).

**Electrical performance of the designed solar cells.** The designed scatterers in this work can enhance the light absorption in the active layer in a broad wavelength region. Here, to investigate the electrical performance of the designed solar cells (Fig. 7), the external quantum efficiency (EQE) and the current-voltage curve are calculated. The results show that the EQEs of solar cells are similar to but smaller than the light absorption efficiencies in active layers due to the influence of carrier recombination. The flat or stepped scatterer can enhance the EQE of the thin-film solar cell in a broad wavelength region for λ>450nm (Fig. 7a). In the short wavelength region of λ<450nm, the main loss is the photo-induced carriers in scatterers and ITO (see Supplementary Note 2). However, the total enhancement of EQE is still remarkable, which can be directly translated into the increases of the short circuit current density ($J_{sc}$). As shown in Fig. 7b, the calculated $J_{sc}$ of the flat solar cell without scatterers is about 11.41 (mA/cm$^2$), which is consistent with the previous experiment[21]. By introducing the above flat or stepped scatterer, $J_{sc}$ can be improved to 13.08 (mA/cm$^2$) and 13.76 (mA/cm$^2$), respectively. Although there is no obvious increase of the open circuit voltage ($V_{oc}$) and the fill factor (FF), the conversion efficiency of the solar cell has a prominent enhancement. The efficiency of 8.1% for a bare solar cell can be improved to 9.37% for the solar cell with a flat scatterer and 9.89% with the stepped scatterer. In particular, the efficiency of the a-Si:H solar cell with the stepped scatterer is close to the world record (10.1%), but it cannot be broken due to the loss of the trapped light energy from the large parasitic absorption in the scatterer in the short wavelength region of λ<450nm.

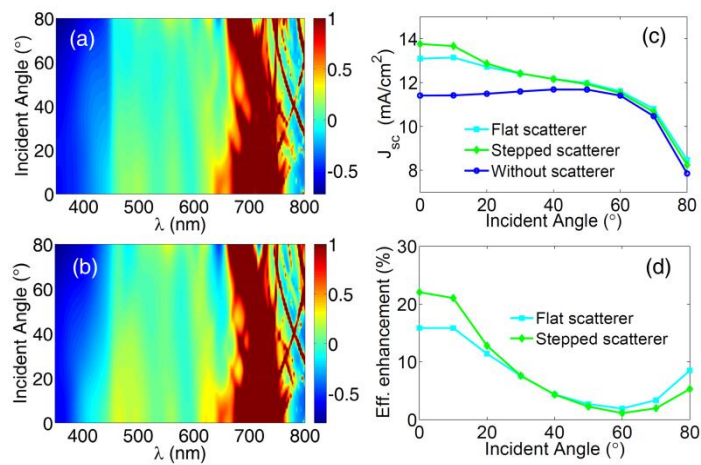

**Figure 8 | Under an oblique incidence, the enhancement of optical and electrical performances for a-Si:H thin-film solar cells with the above designed scatterers.** Light absorption enhancement in the active layers for the solar cell with (**a**) flat and (**b**) stepped scatterers. The values of the enhancements below zero imply the decrease of light absorption, and on the contrary, the values exceeding zero indicate the increase of light absorption. Here the enhancements are truncated by 1 to show clearly the region that has more than twice of the light absorption. (**c**) $J_{sc}$ of the solar cells with and without scatterers under various incident angles. (**d**) The efficiency enhancements of the solar cells with the above designed scatterers as function of the incident angle. The efficiency of bare solar cell is considered as a reference.

Figure 8 shows the angle dependences for the solar cells with the above designed light trapping structures. The enhancements of light absorption indicate that the absorption decreases at λ<450nm that becomes more serious under a large incident angle. Fortunately, the increase of absorption can be sustained for almost all incident angles when λ>450nm (Figs. 8a and 8b). In particular, more than twice of the light absorption in the active layers for about all oblique angles can be gained from λ=680nm to λ=750nm. By collecting the net enhancements of absorption in the solar cells, the improvement of $J_{sc}$ and the efficiency can also be achieved in a wide angle region (Figs. 8c and 8d). But, a remarkable enhancement of $J_{sc}$ induced by the flat and stepped scatterers can only be sustained when the incident angle is within 10°, which decreases rapidly when the oblique angle is greater than this angle. On the contrary, $J_{sc}$ for the bare solar cell increases as the oblique angle increases from 0° to Brewster's angle (around 50°). Taking the efficiency of the bare solar cell as a reference, the enhancement of efficiency of the solar cell with the flat scatterer can reach 15.7% under the normal incident, and that with the stepped scatterer can be up to 22.1%. However, those prominent enhancements can only be maintained within a small angle region (<10°), which decreases rapidly beyond this region till Brewster's angle. Nevertheless, the enhancement of efficiency more than 10% can still be maintained when the incident angle is less than 20°. The reasons behind the above phenomenon are complex, but one of them comes from the increase of parasitic absorptions of the scatterers as the incident angle becomes larger in the short wavelength region (Figs. 8a and 8b). Such parasitic absorption wastes a lot of light energy trapped in the thin-film solar cell.

**Discussion**

In summary, we proposed a novel approach to improve effectively the conversion efficiency of the thin-film solar cells by optimizing the light trapping structures, implemented by combining the GMP method and the full field optical and electrical simulations. The designed light trapping structure is separated from the flat active layer to ensure both the optical benefits and the best electrical performance of the solar cell. The geometrical shape optimization of high-indexed scatterers is used for seeking for the best capture of light in the thin-film solar cell. To implement our strategy, we take the a-Si:H thin-film solar cell as an example, where a flat and a stepped scatterers are designed to provide the excellent light trapping in the cells. With the designed two scatterers, we calculate the maximal enhancement of the conversion efficiency for the a-Si:H thin-film solar cells can be up to 15.7% and 22.1%, respectively, which are better than the previous values[31,34]. Our calculated maximal efficiency (9.89%) of the designed a-Si:H thin-film solar cells is very close to the reported highest value (10.1%). Here we would like to emphasize that our study manifests that there may be two ways to improve further the conversion sufficiency of the thin-film solar cells: the first is to choose the transparent (350nm-800nm) and high indexed (>3) materials as the scatterers with optimized shape structures to reduce the parasitic absorptions in the short wavelength region, and the second is to combine with other light trapping mechanism (such as SPR) to increase the light absorption in the long wavelength region. The strategy proposed in this work is also applicable to the solar cell systems with other materials, which would have great potential to design the thin-film solar cells with high performance.

**Methods**
**Unit cell of the model.** The basic model of the flat a-Si:H thin-film solar cell is adopted according to an actual structure in a previous work[21]. As shown in Figure 1, the single junction a-Si:H solar cell consists of 100nm thick Ag back reflector on $SiO_2$ substrate, 80nm thick AZO layer, 280nm thick active layer (from top to bottom, 10nm thick *p* layer, 250nm *i* layer and 20nm *n* layer ) and 80nm thick ITO layer. Light trapping is based on the high indexed GaP scatterer with height *H* and pitch *L*. The scatterer is placed on the surface of ITO to keep the best electrical performance of the solar cell. $D_2$ symmetry is imposed by to reduce the computational requirements.
**Optical and electrical simulations**. The simulation method to implement the optical and electrical performance is based on the previous work[20,37]. All simulations are carried out by using the finite element method (FEM) software package[37]. The light field distribution in a solar cell can be obtained by solving Maxwell's equations. We assume that the light absorbed by the active layer is all used for exciting carriers. Then, we can use the light field information in the active layer to describe the generation rate of the carriers[20]. Finally, the electrical performance of the solar cell can be gained by calculating the transport equations of carriers and Poisson's equation. The calculation details can be found in Supplementary Note 1.
**Optimization of geometrical shape structure**. In this work, GPM is employed to optimize the geometrical shape of the dielectric scatterer. GPM distributes several control points on the optimization region, and fits a three-dimensional (3D) surface by the heights of those control points, and finally defines the geometrical shape by the intersection curve between a level plane and the 3D surface. By employing the GPM, the shape optimization problem can be transformed into the issue that searches for the optimal heights of the control points. This simplified optimization problem can be readily handled by any optimization algorithm.

In this work, the optimization region is chosen as the cross-section of the unit cell which contains the dielectric scatterer. The shape of the dielectric is defined by the intersection curve between the 3D surface and a level plane. Here, such an intersection curve is approximated by a polygon which has 50~80 sides for accurate calculations. The data information of those sides is converted into the FEM script to realize the connection between FEM and GPM. The shape of the scatterer is changed by adjusting the heights of the control points. The Nelder–Mead algorithm[45] is applied to seek for the optimal heights of those control points in combination with the optical and electrical calculations.

**Acknowledgements**

The authors are benefitted from useful discussions with Q. B. Yan, Z. G. Zhu and Q. R. Zheng. This work is supported in part by the MOST of China (Grant No. 2012CB932900 and No. 2013CB933401), the Strategic Priority Research Program of the Chinese Academy of Sciences (Grant No. XDB07010100), and the China Postdoctoral Science Foundation (2014M550805).

**Author contributions**

D.W. and G. S. conceived the project. D. W. designed and executed the simulations. All authors prepared and contributed to the editing of the manuscript.

# Supplementary Information

## New strategy to promote conversion efficiency using high-index nanostructures in thin-film solar cells

DongLin Wang[1] & Gang Su[1,*]

[1]Theoretical Condensed Matter Physics and Computational Materials Physics Laboratory, School of Physics, University of Chinese Academy of Sciences, P. O. Box 4588, Beijing 100049, China.
*Correspondence and requests for materials should be addressed to G.S. (email: gsu@ucas.ac.cn).

**Supplementary Figures**

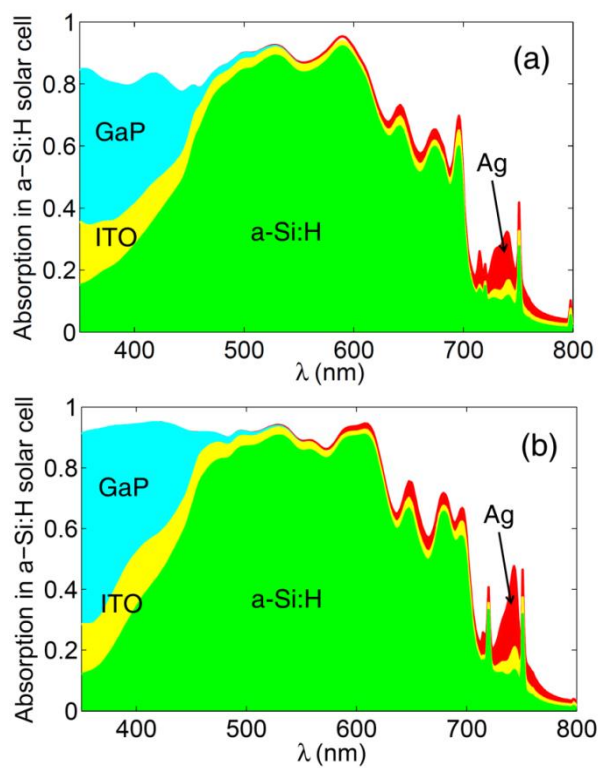

**Supplementary Figure S1. Normalized light absorption in each layers of the a-Si:H thin-film solar cell.** The spectral absorption for the solar cell with the flat scatterer (a) and the stepped scatterer (b). The calculation is carried out under the normal incidence.

**Supplementary Note 1**
**Optical and electrical simulation details**
Based on the FEM software package, Maxwell's equations can be easily solved for executing the optical simulation. All simulations are under a normal incidence except for special instructions. Because the designed scatterer is assumed to have a high symmetry, the transverse electric (TE) and the transverse magnetic (TM) polarized incident light are equivalent under the normal illumination. In this situation, the TE polarized incident light is used for all simulations. Under an oblique incidence, both TE and TM polarized incident light are calculated, and only averaged results are shown. The complex optical constants for Ag, AZO, ITO, a-Si:H and GaP are taken from the previous experimental works[46-49]. The optical absorption within each layers in the solar cell can be obtained by the optical simulation. Mostly, the distribution of electric field intensity ($E(\vec{r},\lambda)$) at each single wavelength is the bridge to connect the optical and electrical simulations. In the electrical simulation, the wavelength dependent generation rate of carriers can be expressed as[19]:

$$G(\vec{r},\lambda) = \frac{\varepsilon'' |E(\vec{r},\lambda)|^2}{2\hbar},$$

where $\varepsilon''$ is the imaginary part of permittivity of a-Si:H. Also, this generation profile should be weighted AM1.5G spectrum[50] to simulate one sun illumination. The details for using the FEM to realize the electrical simulation can be found in a previous report[36]. The only difference here is that more accurate generation profile is used in this work. The electronic material parameters of a-Si:H are taken from the previous studies[51,35]. And 5 $\Omega cm^2$ series resistance and 5 $k\Omega cm^2$ shunt resistance are applied to the solar cell[28].

## Supplementary Note 2
**Parasitic absorption in solar cell**

In the designed solar cell, the light absorption occurs in all layers including scatterers, ITO, AZO, Ag and the active layer. But only the absorption in the active layer can be converted into the photocurrent. The light absorption in other layer can be considered as the parasitic absorption that wastes the light energy trapped by the solar cell. Figure S1 shows the light absorption in all layers of the solar cell with the flat scatterer and the stepped scatterer. Because GaP has a high transmittance in the long wavelength region (>500nm) and a high absorption ability in the short wavelength region (<500nm), a large parasitic absorption is formed in dielectric scatterer when λ<500nm. Besides, the parasitic absorption in ITO and Ag are also the optical loss channels in the solar cell. In this work, the parasitic absorption in GaP scatterers wastes a lot of light energy in the short wavelength region. Choosing another material that has high index and less light absorption may overcome this undesirable optical loss, and could improve further the conversion efficiency of the a-Si:H thin-film solar cell.

## Supplementary References